\documentclass[runningheads]{llncs}
\usepackage{graphicx}
\usepackage{graphicx}
\usepackage{multirow}
\usepackage{tabu}
\usepackage{bbm}
\usepackage{diagbox}
\usepackage[english]{babel}
\usepackage{comment}
\usepackage{array}
\usepackage{amstext}
\usepackage{makecell}
\usepackage{caption}
\usepackage{tablefootnote}
\usepackage{babel}
\usepackage{booktabs} 
\usepackage{color}

\newcommand{\ie}{\textit{i.e.}}
\newcommand{\eg}{\textit{e.g.}}

\begin{document}
\title{Hierarchical Classification of Pulmonary Lesions: A Large-Scale Radio-Pathomics Study}
\titlerunning{Hierarchical Classification of Pulmonary Lesions}

\author{Jiancheng Yang\inst{1,2,3,}\thanks{These authors have contributed equally: Jiancheng Yang and Mingze Gao.}  \and Mingze Gao\inst{3,\star} \and Kaiming Kuang\inst{3} \and Bingbing Ni\inst{1,2,4,}\thanks{Corresponding author: Bingbing Ni.} \and Yunlang She\inst{5} \and Dong Xie\inst{5} \and Chang Chen\inst{5}}

\authorrunning{J. Yang et al.}
%
\institute{Shanghai Jiao Tong University, Shanghai, China\\
	\and MoE Key Lab of Artificial Intelligence, AI Institute, Shanghai Jiao Tong University\\
	\and Dianei Technology, Shanghai, China
	\and Huawei Hisilicon, Shanghai, China\\
	\and Shanghai Pulmonary Hospital, Tongji University, Shanghai, China	\\
	\email{jekyll4168@sjtu.edu.cn}
}

\maketitle              
\begin{abstract}

Diagnosis of pulmonary lesions from computed tomography (CT) is important but challenging for clinical decision making in lung cancer related diseases. Deep learning has achieved great success in computer aided diagnosis (CADx) area for lung cancer, whereas it suffers from label ambiguity due to the difficulty in the radiological diagnosis. Considering that invasive pathological analysis serves as the clinical golden standard of lung cancer diagnosis, in this study, we solve the label ambiguity issue via a large-scale radio-pathomics dataset containing 5,134 radiological CT images with pathologically confirmed labels, including cancers (\eg, invasive / non-invaisve adenocarcinoma, squamous carcinoma) and non-cancer diseases (\eg, tuberculosis, hamartoma). This retrospective dataset, named Pulmonary-RadPath, enables development and validation of accurate deep learning systems to predict invasive pathological labels with a non-invasive procedure, \ie, radiological CT scans. A three-level hierarchical classification system for pulmonary lesions is developed, which covers most diseases in cancer-related diagnosis. We explore several techniques for hierarchical classification on this dataset, and propose a Leaky Dense Hierarchy approach with proven effectiveness in experiments. Our study significantly outperforms prior arts in terms of data scales ($6\times$ larger), disease comprehensiveness and hierarchies. The promising results suggest the potentials to facilitate precision medicine.

\end{abstract}

\section{Introduction}

\begin{figure}[tb]
	\includegraphics[width=\linewidth]{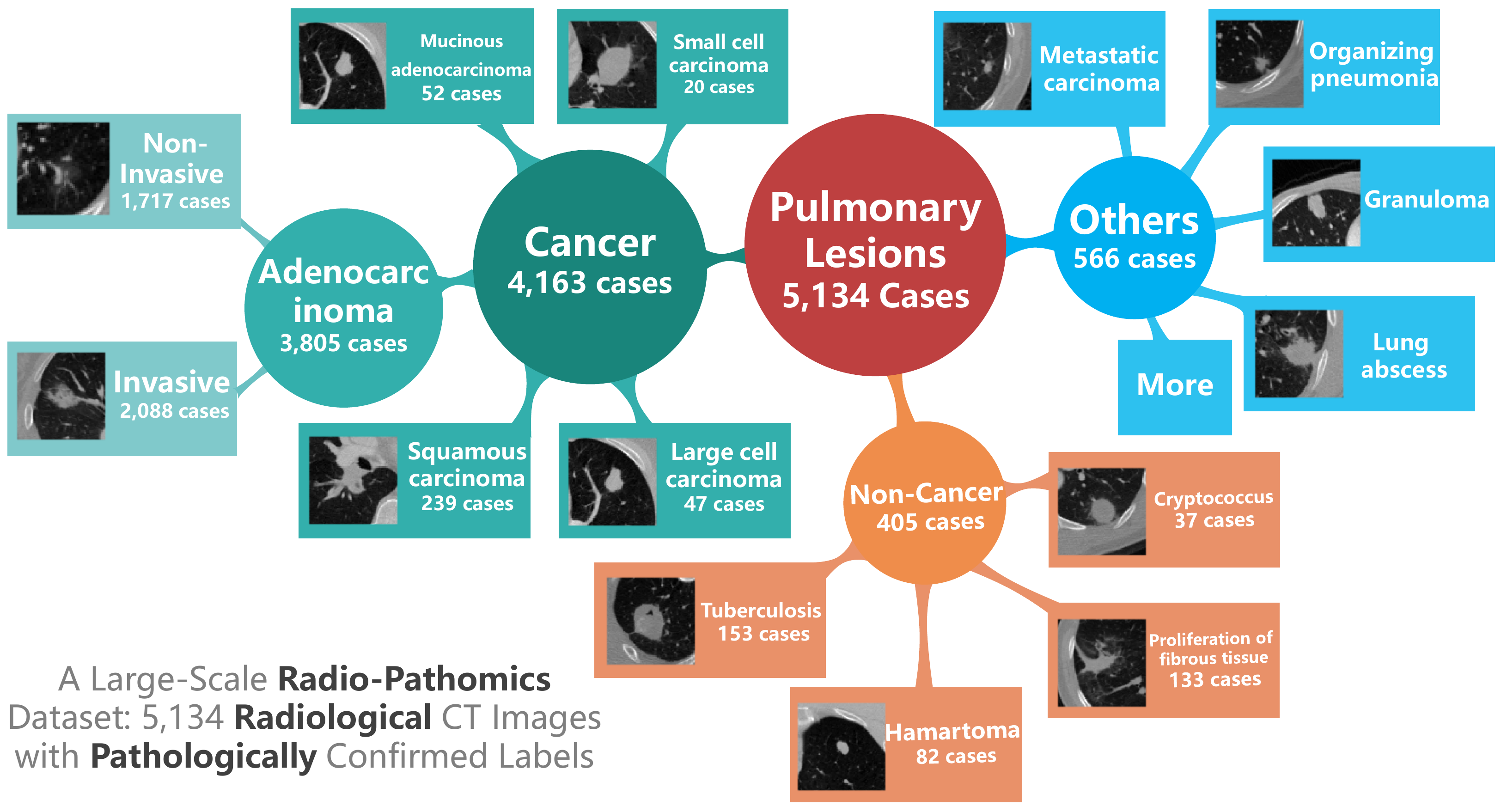}
	\caption{We develop a hierarchical multi-disease classification system of pulmonary lesions based on an in-house large-scale Pulmonary-RadPath dataset, which consists of 5,134 radiological CT cases with pathologically confirmed labels, \ie, the labels are retrospectively collected via invasive pathological analysis. By utilizing this dataset, our hierarchical classification system could predict the \textbf{invasive} pathological labels from \textbf{non-invasive} radiological CT images. Notably, it significantly outperforms prior arts in terms of data scale ($6\times$ larger), disease comprehensiveness and hierarchies. }
	\label{fig:data-overview}
\end{figure}

Lung cancer is the most commonly diagnosed cancer worldwide, which accounts for 18.4\% of global cancer-related mortality in 2018 \cite{Bray2018GlobalCS}. Remarkable success has been achieved in deep learning for lung nodule detection \cite{Dou2017AutomatedPN,Tang2019NoduleNetDF,yang2020relational} and diagnosis \cite{Hussein2017RiskSO,Xie2017TransferableME,yang2019reinventing}, thanks to medical datasets, \eg, LIDC-IDRI \cite{Armato2011TheLI}. However, the radiological diagnosis of lung cancer suffers from ambiguous labels \cite{Yang2019ProbabilisticRA}. For instance, 4 expert annotators make diverse diagnosis in LIDC-IDRI dataset \cite{Armato2011TheLI}. A possible solution to reduce label ambiguity is to leverage clinical golden standard. For lung cancer diagnosis, invasive pathological analysis serves as the golden standard. To this end, we build a large-scale \textbf{radio-pathomics} dataset, consisting of 5,134 \textbf{radiological} CT images with \textbf{pathologically} confirmed labels. Once a deep learning system is well trained on this dataset, it could predict invasive pathological labels via non-invasive procedures, \ie, radiological CT scans. The in-house dataset, named Pulmonary-RadPath, is collected retrospectively from a single clinical center. It covers most diseases in lung cancer-related diagnosis, including cancer (\eg, invasive / non-invaisve adenocarcinoma, squamous carcinoma) and non-cancer diseases (\eg, tuberculosis, hamartoma).

Considering the heterogeneity of included lesions, we develop a three-level hierarchical classification deep learning system of pulmonary lesions with the help of Pulmonary-RadPath dataset. Several hierarchical classification strategies are explored on this real-world clinical dataset, including Leaf-Node baseline approach, Flattened Hierarchy and Leaky Node (details in Sec.~\ref{sec:hier-method}). We further propose (Leaky) Dense Hierarchy approach to encourage hierarchical feature reuse, and prove its effectiveness in our experiments. These strategies improve the three-level hierarchical classification performance compared with naive baseline (Leaf-Node).

\paragraph{\textbf{Contributions.}} We develop a three-level hierarchical classification system for pulmonary lesions, via an in-house large-scale radio-pathomics dataset, where 5,134 radiological CT images with pathologically confirmed labels (clinical golden standard) are collected retrospectively. We explore several hierarchical classification strategies and propose a Leaky Dense Hierarchy approach with proven effectiveness in experiments. Our study significantly outperforms previous studies in terms of data scales ($6\times$ larger), disease comprehensiveness and hierarchies.

\section{Materials and Methods}
\subsection{Pulmonary-RadPath Dataset}

\paragraph{\textbf{Dataset Overview.}}
	Classification of lung cancers is exceedingly complicated due to its complex and diverse nature. According to WHO lung tumor classification guide, there are 77 subtypes of lung tumors \cite{travis20152015}. In this study, we compile a large dataset named Pulmonary-RadPath that contains 5,134 cases with CT scans and invasive pathological diagnosis. All data is collected from a single clinical center (Shanghai Pulmonary Hospital, Tongji University, Shanghai, China). Thicknesses of these scans range between 0.2 mm and 1.0 mm, and the average long diameter of all lesions is 1.9 cm. More than $80\%$ of the cases are diagnosed as malignant. Classes are unbalanced in our dataset, including some diseases with less than 10 cases present. This results from the uneven distribution of lung diseases. 
	
	Since that our research is based on CT scans rather than pathological images, we select some categories as our classification target according to their quantities and visibility in CT volumes. We categorize all individual diseases into three major types: cancer (\eg, adenocarcinoma and squamous carcinoma), non-cancer (\eg, tuberculosis and hamartoma) and others (long tail diseases, \eg, metastatic carcinoma and granuloma). All classes of interest, their tags and corresponding sample sizes in each subset are summarized in Table~\ref{tab:data-overview}. Note that in our study, ``invasive'' denotes	minimally invasive adenocarcinoma (MIA) and invasive pulmonary adenocarcinoma (IA), while ``non-invasive'' denotes atypical adenomatous hyperplasia (AAH) and adenocarcinomas in situ (AIS).

\paragraph{\textbf{Annotation and Pretreatment.}}

{\color{black} 
	
	The hierarchical pathological label and mass center of each lesion is manually labelled by a junior thoracic radiologist, according to corresponding pathological reports. These annotations are then confirmed by a senior radiologist with 15 years of experience in chest CT. Decisions on CT findings are made by consensus. The data is anonymized so that patient identities cannot be traced. We preprocess the data following a common practice \cite{zhao20183d}: (1) Resample CT volumes to $1mm \times 1mm \times 1mm$. (2) Normalize Hounsfield Units to $[-1,1]$. (3) Crop a $48 \times 48 \times 48$ volume centered at the centroid of the each lesion.
	The dataset is randomly split into 5 subsets of same size, numbered as 0 to 4. Each subset retains roughly the same proportion for each disease as the entire dataset. We use subset 0-3 for training and hyperparameter validation, and hold out subset 4 only for testing.

}

\begin{table}[tb]
	\caption{Overview of Pulmonary-RadPath Dataset. There are 5,134 radiological CT cases with pathologically confirmed labels in total, where 80\% are regarded as Train-Dev, and 20\% are regarded as Test. Classification categories are structured into 3 hierarchical levels. Tags are named as ``HeadID+ClassID'' (Fig.~\ref{fig:hierarchy}).}\label{tab:data-overview}
	\centering
	\begin{tabular*}{\hsize}{@{}@{\extracolsep{\fill}}clcccc@{}}
		\toprule
		Level &Class & Tag &  Train-Dev &  Test  & Total\\
		\midrule
		0&ALL &H0&4,109&1,025 & 5,134\\
		1&\hspace{1em}Cancer&H1a & 3,330 & 833 & 4,163 \\
		2&\hspace{2em} Adenocarcinoma&H2a & 3,042 & 763 & 3,805 \\
		3&\hspace{3em} Invasive &H4a& 1,672 & 416 & 2,088 \\
		3&\hspace{3em} Non-invasive &H4b& 1,370 & 3,47 & 1,717 \\
		2&\hspace{2em} Squamous carcinoma &H2b& 192 & 47 & 239 \\
		2&\hspace{2em} Large cell carcinoma &H2c& 38 & 9 & 47 \\
		2&\hspace{2em} Small cell carcinoma &H2d& 16 & 4 & 20 \\
		2&\hspace{2em} Mucinous adenocarcinoma & H2e& 42 & 10 & 52 \\
		1&\hspace{1em} Non-Cancer&H1b& 326 & 79 & 405 \\
		2&\hspace{2em} Tuberculosis &H3a& 123 & 30 & 153 \\
		2&\hspace{2em} Hamartoma &H3b& 66 & 16 & 82 \\
		2&\hspace{2em} Proliferation of fibrous tissue &H3c& 107 & 26 & 133 \\
		2&\hspace{2em} Cryptococcus &H3d& 30 & 7 & 37 \\
		1&\hspace{1em}Others &H1c& 453 & 113 & 566 \\
		\bottomrule
	\end{tabular*}
	
\end{table}

\subsection{Approaches for Hierarchical Classification} \label{sec:hier-method}

\begin{figure}[tb]
	\includegraphics[width=\linewidth]{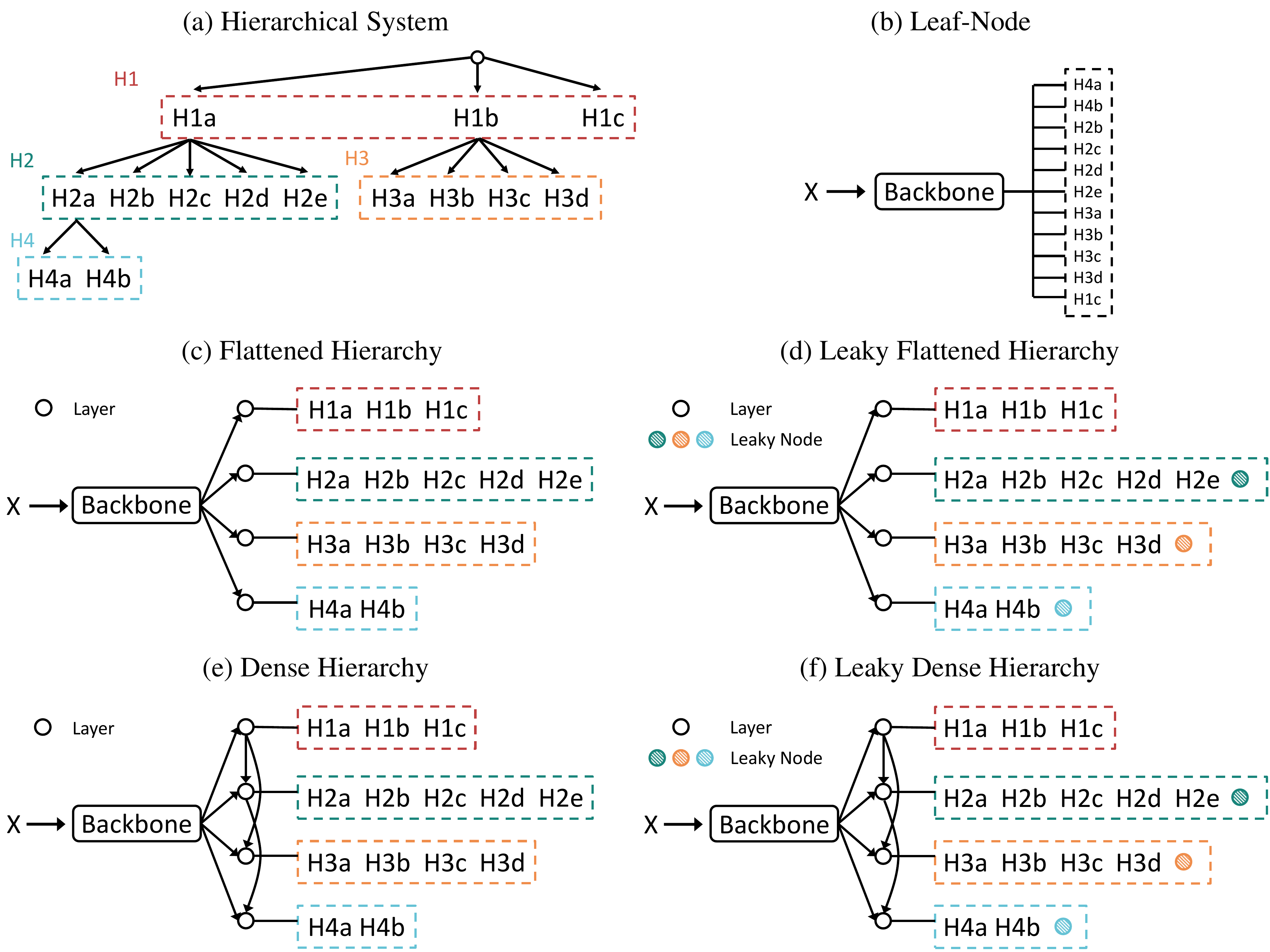}
	\caption{Approaches for hierarchical classification in this study. (a) The hierarchical classification system for pulmonary lesions. The tags in the tree follow the names from Table \ref{tab:data-overview}. We use a same 3D DenseNet backbone \cite{huang2017densely} for all our experiments. (b) Leaf-Node approach. The leaf nodes are ordered from left to right. (c) Flattened Hierarchy approach. (d) Leaky Flattened Hierarchy approach. (e) Dense Hierarchy approach. (f) Leaky Dense Hierarchy approach.} \label{fig:hierarchy}
\end{figure}

\paragraph{\textbf{Model Overview.}}

{\color{black} 
	
We use a custom 3D DenseNet \cite{huang2017densely} as the backbone. DenseNet has shown compelling accuracies on natural image recognizing tasks with efficient use of parameters. To leverage the power of dense connectivities, we transform the 2D DenseNet into its 3D variant. The 3D DenseNet comprises of stacked densely-connected blocks (\ie, DenseBlock), each of which consists of several convolutional modules. In each convolutional module, a $1 \times 1 \times 1$ convolution layer with 64 filters is followed by a $3 \times 3 \times 3$ convolution layer with 16 filters, Dense connections enable efficient 3D representation learning. In this study, we use a 3D DenseNet that has a growth rate of 16 and three DenseBlocks with depths of 8, 4 and 4, respectively. The features after global average pooling layer are then processed by either of the following hierarchical classification heads, \ie, (b) (c) (d) (e) (f) in Fig. \ref{fig:hierarchy}.

We arrange the hierarchical system of disease categories in a tree structure, as shown in Fig.~\ref{fig:hierarchy}~(a). Three root nodes, H1a, H1b and H1c, represent general disease categories: cancer, non-cancer and others. Individual diseases are initialized as leaf nodes. The entire tree is then developed in a bottom-up fashion. Nodes under the same pathological class are merged as internal nodes, until all classes and sub-classes are included in the tree.
	
}

\paragraph{\textbf{Leaf-Node Approach.}}

We first introduce the naive baseline Leaf-Node approach~\cite{Esteva2017DermatologistlevelCO} for hierarchical classification. Leaf-Node approach treats hierarchical classification as a simple multi-class problem, where each leaf node in the hierarchical system is a class. The model outputs the probabilitiy for each individual diseases. The probabilities of internal or root nodes equal to the sum of those of all the children nodes.

\paragraph{\textbf{Flattened Hierarchy Approach.}}

Inspired by the multiple softmax heads for individual annotators \cite{Guan2017WhoSW}, we propose the Flattened Hierarchy approach with multiple heads for different hierarchies, \ie, H1, H2, H3 and H4 shown in Fig.~\ref{fig:hierarchy}~(a). Instead of structuring it as a plain multi-class problem, the Flattened Hierarchy approach includes each training sample's hierarchical information, \ie, all categories on the path from itself to the top node. The problem is then decomposed into several multi-class classifications within each hierarchy. Probabilities of each leaf node is calculated by multiplying those of nodes on the path from the leaf node to the root. Compared with the Leaf-Node approach, Flattened Hierarchy enables the model to learn the hierarchical relations between classes. Loss is ignored for a certain hierarchy if the target class is not in the hierarchy.

\paragraph{\textbf{Dense Hierarchy Approach.}}

To reuse features in different hierarchies, we propose the Dense Hierarchy approach as shown in Fig.~\ref{fig:hierarchy}~(e). Directed connections are made between any parent and child hierarchy, where features from the parent hierarchy are passed down to the next level and concatenated with features of child hierarchy. This design is similar to DenseNet \cite{huang2017densely}, in that features of the higher hierarchy are reused at a lower level through jump connections. Loss is ignored for a certain hierarchy if the target class is not in the hierarchy.

\paragraph{\textbf{Leaky Node as a Virtual Class.}}

We observe better performances compared with the Leaf-Node approach when adopting strategies above. Moreover, we introduce a virtual node inspired by hierarchical novelty detection \cite{Lee2018HierarchicalND}, named Leaky Node, to further improve the performance in hierarchical classification settings. This technique could be easily applied to both Flattened and Dense Hierarchy approaches as a plug-and-play unit, and it stably delivers better performances in our experiments. The Leaky Node is a fall-back class when a certain training sample does not belong to any nodes in this hierarchy. In such cases, the leaky node label is set positive while all others are negative, and its loss is back-propagated along with those of other nodes in the same hierarchy. The leaky nodes make the full data distributions accessible to the hierarchical head classifiers; as the softmax classification with cross-entropy training loss is naturally proper scoring rules~\cite{Gneiting2007StrictlyPS}, the leaky node approach encourages calibration \cite{Guo2017OnCO} of predictive uncertainty in each hierarchy classifier, which leads to better classification performance in our experiments (Sec. \ref{sec:result}).

\subsection{Model Training and Inference}

{\color{black} 
	
	During training, we use cross entropy loss with class-imbalance weights allocated for each head. Online data augmentations, including random rotation, flipping and translation are applied at each volume. We use Adam optimizer \cite{Kingma2014AdamAM} to train all models end-to-end for 200 epochs. We set the initial learning rate at 0.01, and decay it by a factor of $1/3$ every 20 epochs. The batch size is 16 for training. We use synchronized batch normalization \cite{Ioffe2015BatchNA,Zhang_2018_CVPR} for multiple GPUs. All experiments are implemented with PyTorch 1.2 \cite{Paszke2017AutomaticDI} and 4 NVIDIA 1080Ti GPUs. 
	
	For the Leaf-Node approach, we calculate an internal or root node's probability by summing up those of all its children. In terms of Hierarchy approach, the probability of leaf or internal node is calculated by multiplying those of nodes on the path from itself to the root.

}

\section{Experiments}

\subsection{Evaluation}

We calculate the Area Under the Receiver Operating Characteristic Curve (AUROC) for each class to assess the agreement between the ground truth and the model output. Within each head, the sample's class is labelled as 1 and the others are set to 0. We calculate each head's mAUC to evaluate each hierarchical level performance. The mAUC is mean AUC weighted by number of samples in each class. The mAUC@L is calculated by averaging AUCs of all leaf nodes, weighted by number of samples at each leaf node. Similarly, the mAUC@H1-4 is the average of AUCs at all child nodes within the hierarchy.

\subsection{Results} \label{sec:result}

\begin{table}[tb]
	\caption{Mean AUC (mAUC) within each head (H1/2/3/4) and leaf nodes (L). Performance of each node is depicted in Table~\ref{tab:head-performance}.}\label{tab:mAUC}
	\centering
	\begin{tabular*}{\hsize}{@{}@{\extracolsep{\fill}}lccccc@{}}
		\toprule
		Methods & mAUC@H1 & mAUC@H2 &  mAUC@H3 & mAUC@H4 & mAUC@L\\
		\midrule
		Leaf-Node & 68.9 & 86.9 & 76.5 & 43.8 & 80.7\\
		Flattened Hierarchy & 64.3 & 85.5 & 76.2 & 92.2 & 80.4\\
		Leaky Flattened Hierarchy & 74.1 & 87.3 & 70.0 & 92.4 & 81.9\\
		Dense Hierarchy & 66.1 & 87.3 & \textbf{80.0} & 93.0 & 81.1\\
		Leaky Dense Hierarchy & \textbf{75.9} & \textbf{89.1} & 79.9 & \textbf{93.8} & \textbf{84.1}\\
		
		\bottomrule
	\end{tabular*}
	
\end{table}

\begin{table}[tb]
	
	\caption{Model performance in AUC of each node class within H1/2/3/4. AUC is computed between each node class and every other classes in a same node.}\label{tab:head-performance}
	\centering
	\scriptsize
	\begin{tabular*}{\hsize}{@{}@{\extracolsep{\fill}}l|ccc|ccccc|cccc|cc@{}}
		\toprule
		Methods &H1a & H1b & H1c & H2a & H2b& H2c & H2d & H2e & H3a & H3b & H3c & H3d & H4a & H4b  \\
		\midrule
		Leaf-Node & 69.3 & 67.8 & 66.6 & 87.1 & 86.8 & 86.2 & 76.4 & 74.6 & 70.6 & 71.6 & 88.0 & \textbf{69.8} & 55.3 &31.0\\
		Flattened Hierarchy & 64.4 & 62.9 & 64.3 & 85.7 & 87.2 & 84.8 & 78.3 & 68.2 & 76.2 & 61.9 & 88.2 & 63.9 & \multicolumn{2}{c}{92.2}\\
		Leaky Flattened Hier. & 74.6 & 72.9 & 71.1 & 87.4 & 86.2 & \textbf{92.8} & 80.3 & \textbf{81.9} & 76.6 & \textbf{78.3} & 88.8 & 54.2 & \multicolumn{2}{c}{92.4}\\
		Dense Hierarchy & 66.2 & 67.2 & 64.6 & 87.5 & 87.2 & 89.0 & \textbf{83.8} & 75.0 & 77.8 & 69.9 & \textbf{91.5} & 69.6 & \multicolumn{2}{c}{93.0}\\
		Leaky Dense Hierarchy & \textbf{76.4} & \textbf{76.9} & \textbf{71.5} & \textbf{89.2} & \textbf{90.4} & 86.9 & \textbf{83.8} & 75.9 & \textbf{79.5} & 71.1 & 90.2 & 63.3 & \multicolumn{2}{c}{\textbf{93.8}}\\
		
		\bottomrule
	\end{tabular*}
	
\end{table}

	Table~\ref{tab:mAUC} shows the mAUC of each head and the mAUC over all leaf nodes. Table~\ref{tab:head-performance} gives the details of each node's AUC. At each level, at least one of our hierarchical methods achieves better AUC score than the Leaf-Node approach. Leaf-Node does not even delievers the best performance on mAUC@L, which only includes AUC scores on leaf nodes. This is because the Leaf-Node approach does not include the hierarchical relation and the differences in the level of difficulty between each classification task. For instance, Leaf-Node gets the lowest mAUC in H4 even though this hierarchy makes up more than $70\%$ of total data.
	
	It is worth noting that Dense Hierarchy approaches reach higher mAUC than its Flattened counterparts in all hierarchies, as they share features through jump connections between levels. Besides, leaky nodes consistently improve the network's performance, which demonstrates the usefulness of learning calibrated classification as real-world or full data distribution.

\section{Discussion}

\paragraph{\textbf{Compared with Previous Studies.}}

In this study, we develop a radio-pathomics hierarchical classification system of pulmonary lesions based on 3D deep learning. The labels are retrospectively collected via invasive pathological analysis. Although the labels are collected invasively, once the deep models are trained on this dataset, we could predict the invasive labels via a non-invasive procedure (\ie, radiological CT scans). Note that the clinical golden standard pathological labels avoid the annotation ambiguity \cite{Yang2019ProbabilisticRA} from human annotators. Even for experts, the diagnosis could be diverse for controversial cases (refer to the diverse annotations from 4 experts in LIDC-IDRI dataset \cite{Armato2011TheLI}).  \emph{DenseSharp} \cite{zhao20183d} introduces deep learning methodology in the radio-pathomics research for lung cancer. A dataset of 653 adenocarcinoma CT scans is built to develop and validate the deep learning system. It achieves a promising AUC performance of 78.8 in differentiating invasive and non-invasive adenocarcinoma, which outperforms radiologists in the observer study. This study focuses on adenocarcinoma; though clinically important, it is only a possible diagnosis in the pulmonary lesion diagnosis system, which hinders the clinical usage in real world. Several studies \cite{gong2019deep,kim2020ct} follows the same setting as \emph{DenseSharp} \cite{zhao20183d} with 828 and 525 cases, which results in AUC of 92.0 and 92.1, respectively. As a comparison, the classification system of our study is more comprehensive in terms of disease coverage and hierarchical levels. We do not only include adenocarcinoma, but also squamous carcinoma and other cancer and non-cancer diseases. On our dataset, we achieve a AUC performance of 93.8 in differentiating invasive and non-invasive adenocarcinoma\footnote{Note that the AUC metrics could not directly compare with each other since the datasets, inclusion criteria, and experiment settings are different.}. We believe the performance comes from the overwhelming superiority over previous studies on data scale, disease comprehensiveness and hierarchies.

\paragraph{\textbf{Data Bias and Long-Tailed Issues.}} A limitation of this study is the data bias and long-tailed issues for subjects with pathological analysis. Pathological analysis as a invasive or minimally invasive examination, is generally performed on subjects regarded as high-risk by clinicians. It is interesting to consider how to calibrate the trained models with pathologically confirmed labels to real-world distribution \cite{Guo2017OnCO}. Besides, cancers (\eg, squamous carcinoma) not suitable for surgery are also lack of pathological analysis. On the other hand, certain types of cancer or diseases are naturally very long-tailed, \eg, small cell carcinoma. Techniques for long-tailed classification \cite{Cui2019ClassBalancedLB} or even few-shot classification \cite{Finn2017ModelAgnosticMF} are worth explorating the hierarchical settings. It is also beneficial to integrate model uncertainty \cite{huang2019evaluating} into the deep hierachical classification system and improve model generalization~\cite{zhao2019convolution} upon diverse imaging protocols.

\section{Conclusion}

In this study, we develop a hierarchical multi-disease classification system of pulmonary lesions with on an in-house large-scale radio-pathomics dataset, named Pulmonary-RadPath. Several hierarchical classification strategies are explored and developed on our dataset. Our study significantly outperforms prior arts in terms of data scale, disease comprehensiveness and hierarchies. 

In future studies, we will continuously complement the Pulmonary-RadPath data with more classes and more samples (especially the highly imbalanced classes). Research is urged on the data bias and long-tailed issues. It is also interesting to integrate radio-genomics (\eg, predicting EGFR-mutation \cite{Wang2019PredictingEM,Zhao2019TowardAP}) into our radio-pathomics hierarchical classification system.

\subsubsection{Acknowledgment.}
This work was supported by National Science Foundation of China (61976137, U1611461). This work was also supported by Shanghai Municipal Health Commission (2018ZHYL0102, 2019SY072, 201940018). Authors appreciate the Student Innovation Center of SJTU for providing GPUs.

\bibliographystyle{splncs04}

\end{document}